\documentclass[twocolumn,pra,amsmath,amssymb,floatfix,superscriptaddress]{revtex4-1}
\usepackage{graphicx}
\usepackage{dcolumn}
\usepackage{bm}
\usepackage{xspace}
\usepackage{hyperref}
\usepackage{color}
\def\new{\color{black}}
\providecommand*{\et}{\emph{et\,al.}\xspace}%
\providecommand*{\tc}{$T_c$\xspace}%
\providecommand*{\kfs}{K$_x$Fe$_{2-y}$Se$_2$\xspace}%
\begin{document}

\title{Suppression of the antiferromagnetic order when approaching the superconducting state in {\new a phase-separated crystal of} K$_x$Fe$_{2-y}$Se$_2$}
\author{Shichao~Li}
\author{Yuan~Gan}
\author{Jinghui~Wang}
\affiliation{National Laboratory of Solid State Microstructures and Department of Physics, Nanjing University, Nanjing 210093, China}
\author{Ruidan~Zhong}
\affiliation{Condensed Matter Physics and Materials Science
Department, Brookhaven National Laboratory, Upton, New York 11973,
USA}
\affiliation{Materials Science and Engineering Department, Stony Brook University, Stony Brook, New York 11794, USA}
\author{J.~A.~Schneeloch}
\affiliation{Condensed Matter Physics and Materials Science
Department, Brookhaven National Laboratory, Upton, New York 11973,
USA}
\affiliation{Department of Physics and Astronomy, Stony Brook University, Stony Brook, New York 11794, USA}
\author{Zhijun~Xu}
\affiliation{Department of Physics, University of California, Berkeley, California 94720, USA.}
\affiliation{Materials Science Division, Lawrence Berkeley National Laboratory, Berkeley, California 94720, USA}
\author{Wei~Tian}
\author{M.~B.~Stone}
\author{Songxue~Chi}
\author{M.~Matsuda}
\affiliation{Quantum Condensed Matter Division, Oak Ridge National Laboratory, Oak Ridge,
Tennessee 37831, USA.}
\author{Y.~Sidis}
\author{Ph.~Bourges}
\affiliation{Laboratoire L\'eon Brillouin, CEA-CNRS, Universit\'{e} Paris-Saclay, CEA Saclay, 91191 Gif-sur-Yvette, France}
\author{Qiang~Li}
\author{Genda~Gu}
\author{J.~M.~Tranquada}
\author{Guangyong Xu}
\affiliation{Condensed Matter Physics and Materials Science
Department, Brookhaven National Laboratory, Upton, New York 11973, USA}
\author{R. J. Birgeneau}
\affiliation{Department of Physics, University of California, Berkeley, California 94720, USA.}
\affiliation{Materials Science Division, Lawrence Berkeley National Laboratory, Berkeley, California 94720, USA}
\affiliation{Department of Materials Science and Engineering, University of California, Berkeley, California 94720, USA.}
\author{Jinsheng~Wen}
\email{jwen@nju.edu.cn}
\affiliation{National Laboratory of Solid State Microstructures and Department of Physics, Nanjing University, Nanjing 210093, China}
\affiliation{Collaborative Innovation Center of Advanced Microstructures, Nanjing University, Nanjing 210093, China}


\begin{abstract}
We have combined elastic and inelastic neutron scattering techniques, magnetic susceptibility and resistivity measurements to study single-crystal samples of \kfs, {\new which contain the superconducting phase that has a transition temperature of $\sim$31~K}. In the inelastic neutron scattering measurements, we observe both the spin-wave excitations resulting from the block antiferromagnetic ordered phase and the resonance that is associated with the superconductivity in the superconducting phase, demonstrating {\new the coexistence of these two orders}. From the temperature dependence of the intensity of the magnetic Bragg peaks, we find that well before entering the superconducting state, the development of the magnetic order is interrupted, at $\sim$42~K. {\new We consider this result to be evidence for the physical separation of the antiferromagnetic and superconducting phases; the suppression is possibly due to the proximity effect of the superconducting fluctuations on the antiferromagnetic order.}
\end{abstract}
\maketitle

\section{Introduction}
The A$_x$Fe$_{2-y}$Se$_2$ (A = alkaline metal) superconductors with critical temperature \tc above 30~K, have been studied extensively.\cite{scinkfese} Besides the high \tc, they exhibit several exotic physical properties distinguishing them from other Fe-based systems.\cite{RevModPhys.85.849} First, unlike other systems where there are hole pockets near the Brillouin zone center (the $\Gamma$ point), and electron pockets near the zone corner (the M point),\cite{dinghongepl} in A$_x$Fe$_{2-y}$Se$_2$, the hole band at the $\Gamma$ point sinks well below the Fermi level.\cite{2011arXiv1101.4556M,hongdingarpeskfese,kfesearpes1} This seriously challenges the $s_\pm$-wave pairing symmetry scenario where the interband pairing occurs between the hole and electron bands at the $\Gamma$ and M points, respectively.\cite{Physics.4.26} Second, the high antiferromagnetic ordering temperature {\new ($T_N\sim$560~K)} and large ordered moment {\new ($\sim$3.3~$\mu_{\rm B}$)} have far exceeded those of other Fe-based superconductor systems.\cite{2011arXiv1102.0830B} {\new Below $T_N$, the magnetic peak intensity increases monotonically with decreasing temperature.\cite{2011arXiv1102.0830B,2011arXiv1102.2882Y} This magnetic phase is believed to result from the ordering of the Fe vacancies, which occurs at $T_s$, about 20~K above $T_N$.\cite{2011arXiv1102.0830B} At $T_s$, the sample undergoes a transition from the $I4/mmm$ to the $I4/m$ phase upon cooling.\cite{2011arXiv1102.0830B}} More surprisingly, some early studies have suggested that such a strong magnetic order coexists with the high-\tc superconductivity microscopically.\cite{BAOWei:027402,2011arXiv1101.1873S} However, there is substantial evidence from various techniques that the antiferromagnetic and superconducting orders occur in {\new separated regions of the sample} with the preferred state being controlled by the local concentration.\cite{PhysRevB.83.140505,2011arXiv1106.3026C,nc_4_1897,2011arXiv1107.0412R,phasesep122_1,2011arXiv1108.0069L,PhysRevB.93.075155}
{\new Specifically, the magnetic order occurs in the regions with $x=0.8$ and $y=0.4$, while superconductivity resides   
in the regions with $x\sim1$ and $y=0$.\cite{PhysRevB.83.140505,2011arXiv1106.3026C,nc_4_1897,2011arXiv1107.0412R,phasesep122_1,2011arXiv1108.0069L,PhysRevB.93.075155}
} 

In the BaFe$_2$As$_2$ system, it is commonly believed that the antiferromagnetic order and superconductivity coexist microscopically, and when the system is cooled below \tc, the density of states contributing to the magnetic order is reduced, since some electrons are gapped. {\new The consequence is that the magnetic order parameter is suppressed at \tc, observable as a pronounced reduction of the magnetic peak intensity in elastic neutron scattering experiments.\cite{PhysRevLett.103.087001,PhysRevB.83.172503,PhysRevLett.110.147003,PhysRevB.81.134512} However, for K$_x$Fe$_{2-y}$Se$_2$ samples with the average $y$ much greater than 0 but also show superconductivity, the evidence is less clear,\cite{2011arXiv1102.2882Y,2011arXiv1102.0830B}} because superconductivity only occurs in a small part of the whole sample, that with $y\sim$0, which is surrounded by a strong antiferromagnetic environment with $y\sim$0.4.\cite{scinkfese,PhysRevLett.109.267003} In this case, antiferromagnetism dominates, and superconductivity hardly affects it.

\begin{figure}[htb]
\includegraphics[width=0.95\linewidth,trim=30mm 45mm 10mm 65mm,clip]{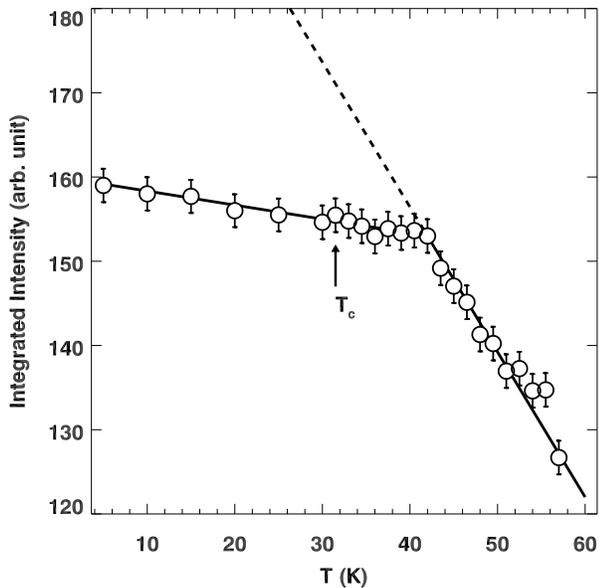}
\caption{Integrated intensity of the scans along the [120] direction through the magnetic Bragg peak (0.2,\,0.4,\,0.5) [scans are shown in Fig.~\ref{scan}(a)]. Lines through data are guides to the eye. The dashed line is the extrapolation to the high-temperature data. Throughout the paper, error bars represent one standard deviation. {\new The data plotted here were collected on HB1A, and were confirmed on HB1.}}
\label{order}
\end{figure}

In order to investigate the interplay between these two orders, it is necessary that a reasonably large portion of the sample has $y\sim$0 and therefore falls in the superconducting phase.\cite{scinkfese,PhysRevB.93.075155} We have obtained such samples,\cite{interplaywen,PhysRevB.86.144530,2011arXiv1111.1026T,2011arXiv1110.5529H} and performed magnetic susceptibility, resistivity, elastic and inelastic neutron scattering measurements on them {\new in the temperature range relevant to the superconductivity}. Our key findings are presented in Fig.~\ref{order}, which shows a clear and unambiguous kink in the magnetic order parameter, occurring at a temperature of $T=42$~K, about 11~K higher than \tc. {\new We believe that this is most likely due to the proximity effect of the superconducting fluctuations on the antiferromagnetic order in the phase-separated sample.\cite{2011arXiv1111.1860J}}

\begin{figure*}[htb]
\includegraphics[width=0.95\linewidth,trim=30mm 45mm 5mm 80mm,clip]{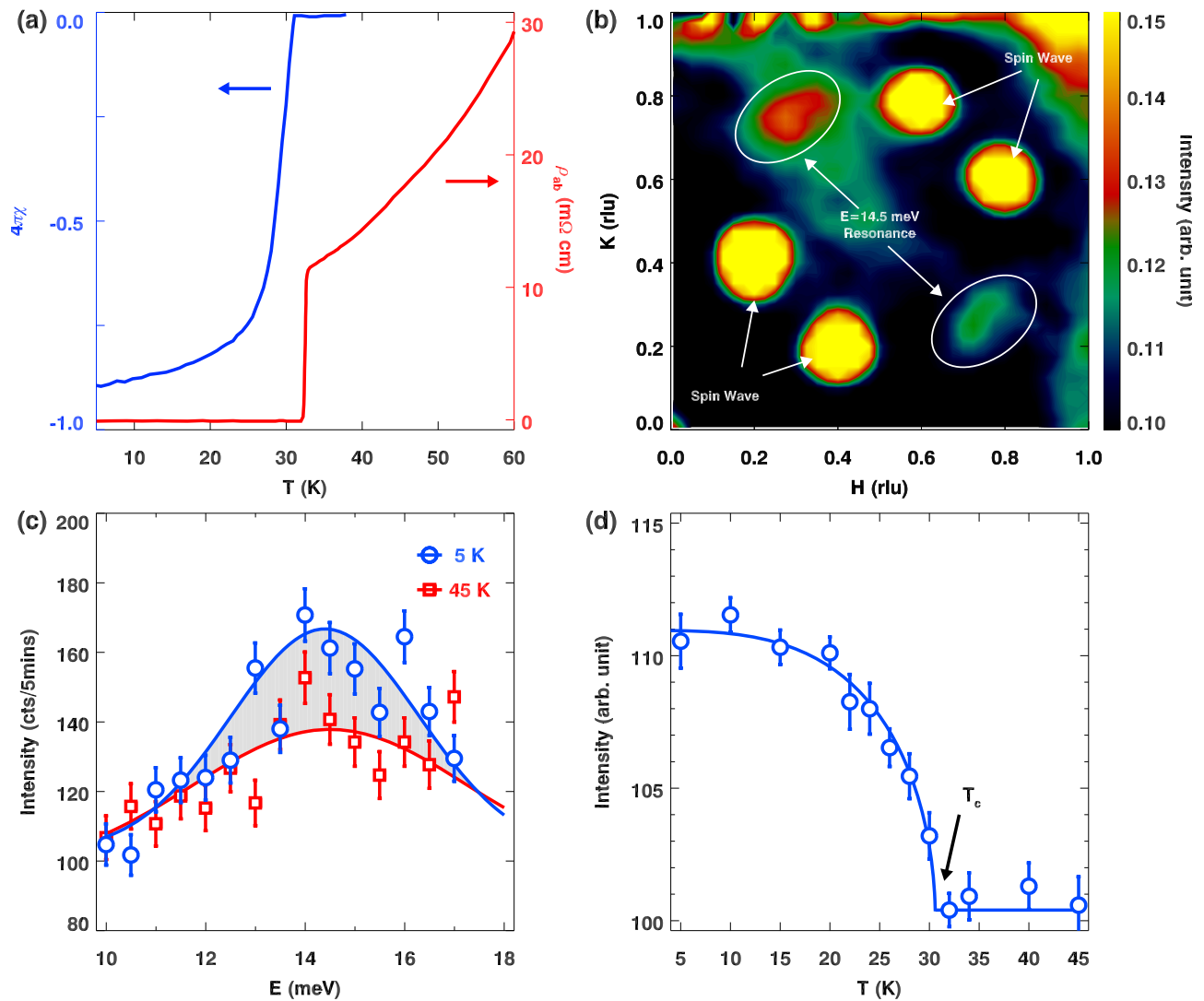}
\caption{(a) Left axis: magnetic susceptibility measured under zero-field-cooling conditions with a magnetic field of 10~Oe applied along $c$ axis. Right axis: resistivity measured in the $a$-$b$ plane in zero field. (b) Contour map of the time-of-flight data projected onto the ($H,\,K,\,0$) plane at 5~K with energies ranging from 12 to 16~mev. The positions of the spin-wave excitations and the resonance mode have been marked. (c) Energy scans at {\new $\bm{Q}=(0.25,\,0.75,\,0)$} at 5 and 45~K. {\new These are the raw data without subtracting the backgrounds.} We counted 15 minutes for each data point and normalized the counts to 5 minutes per point. Lines through data are fits using Lorentzian functions. The shade illustrates the intensity gain at 5~K. (d) Integrated intensities obtained from the fits to the $\bm{Q}$ scans at 14.5~meV through {\new (0.25,\,0.75,\,0)} along the [110] direction. The solid line is a fit to the data using the BCS gap function. {\new The results in (b) were obtained on ARCS, and those in (c) and (d) were obtained on HB1 and confirmed on 1T.}}
\label{ins}
\end{figure*}

\begin{figure*}[htb]
\includegraphics[width=0.98\linewidth,trim=25mm 55mm 10mm 165mm,clip]{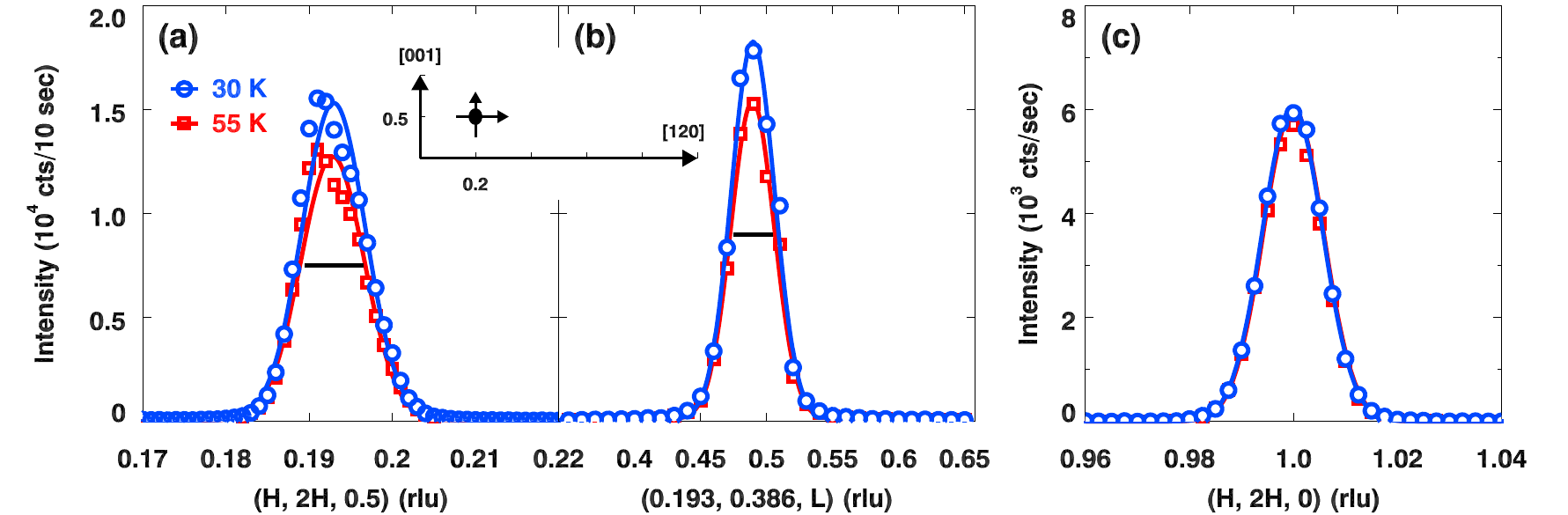}
\caption{(a) and (b), linear scans through the magnetic Bragg peak (0.2,\,0.4,\,0.5) along the [120] and [001] directions, respectively. Lines through data are fits with Gaussian functions. Scan trajectories are shown in the inset. Horizontal bars illustrate the instrumental resolutions. (c) Scans through the nuclear Bragg peak (120) at 30 and 55~K. {\new The data plotted here were collected on HB1A, and were confirmed on HB1.}}
\label{scan}
\end{figure*}

\section{Experimental}
Single-crystal samples of K$_{x}$Fe$_{2-y}$Se$_{2}$ were grown by the self-flux method as described in ref.~\onlinecite{interplaywen}. Magnetic susceptibility, resistivity and neutron scattering measurements were performed on the single-crystal pieces extracted from the same batch. Susceptibility and resistivity were measured using a Quantum Design Magnetic Properties Measurement System and a Physical Properties Measurement System. For the neutron scattering experiments, we used a 5-g single crystal. {\new For this crystal, the exact ratio between the antiferromagnetic (K$_{0.8}$Fe$_{1.6}$Se$_2$) and superconducting (KFe$_2$Se$_2$) phases are not known,\cite{PhysRevB.83.140505,2011arXiv1106.3026C,nc_4_1897,2011arXiv1107.0412R,phasesep122_1,2011arXiv1108.0069L,PhysRevB.93.075155} and we therefore label the sample as \kfs. Nevertheless, as we demonstrate below, we have a reasonably large portion that is superconducting and the superconductivity shows clear effect on the antiferromagnetic phase.} Elastic neutron scattering measurements were carried out on triple-axis spectrometers (TAS) HB1A (incident energy $E_i=14.7$~meV) and HB1 (final energy $E_f=14.7$~meV) located at the High Flux Isotope Reactor (HFIR) in the ($H,\,2H,\,L$) plane. Inelastic neutron scattering experiments were performed on TAS HB1 located at HFIR and 1T located at Laboratoire L\'eon Brillouin (CEA-Saclay), as well as on a time-of-flight (TOF) spectrometer ARCS located at the Spallation Neutron Source (SNS). For the inelastic measurements on HB1 and 1T, data were collected in the ($H,\,K,\,0$) plane with a fixed final energy $E_f=14.7$~meV. On ARCS, we chose an $E_{i}$ of 35~meV. The sample was aligned such that the [001] direction was along the incident beam, and the [110] direction was along the vertical direction. The sample for the neutron scattering experiments had a mosaic spread of 1 degree. {\new Because the sample is air sensitive, we always handled the sample inside a glovebox filled with inert gas. After visual check of the sample, we sealed it into an aluminum can filled with He gas. A leakage check was performed to assure a good seal. In each neutron scattering experiment, we did alignment scans to make sure that the sample was still intact, {\it e.g.}, by examining the intensities of the nuclear Bragg peaks, and the physical positions of the reflection planes. Moreover, we obtained reasonably strong signals in both the elastic and inelastic measurements, confirming the quality of the sample. After each experiment, the sample and the can were stored in a glovebox as a whole. The neutron scattering data are described in reciprocal lattice units (rlu) of ($a^{*},\,b^{*},\,c^{*}$) = $(2\pi/a,\,2\pi/b,\,2\pi/c)$, where $a=b\approx3.88$~\AA, and $c\approx7.21$~{\AA} at room temperature in the $I4/m$ notation.}

\section{Results}
We have measured both the susceptibility and resistivity for several single-crystal pieces of K$_{x}$Fe$_{2-y}$Se$_{2}$, and the results are shown in Fig.~\ref{ins}(a). The $T_c$ is determined to be 31~K from the onset of the diamagnetism. The shielding volume fraction is close to 100\%, {\new estimated after correcting the geometrical factor for the rectangular-bar-shape sample. The resistivity reaches zero at 32~K, slightly higher than the $T_c$, which may be an indication of a very slight sample inhomogeneity.} In Fig.~\ref{ins}b, we plot the TOF data measured at 5~K, with energies integrated from 12 to 16~meV. Intensities are averaged over the full range of $\bm{Q_z}$. Clearly, there are two sets of signals {\new in the two-dimensional Brillouin zone}, specifically, one around (0.5${\pm}$0.3, 0.5${\pm}$0.1) and (0.5${\pm}$0.1, 0.5${\pm}$0.3) which represents the spin-wave excitations originating from the $\sqrt{5}\times\sqrt{5}$ block antiferromagnetic order,\cite{2011arXiv1105.4675W} and another one around (0.5${\pm}$0.25, 0.5${\mp}$0.25) which represents the neutron-spin resonance mode at 14.5~meV.\cite{resonancekfese} {\new The former originates from the antiferromagnetic phase with $y\sim$0.4 and the latter from the superconducting phase with $y\sim$0.\cite{PhysRevB.83.140505,2011arXiv1106.3026C,nc_4_1897,2011arXiv1107.0412R,phasesep122_1,2011arXiv1108.0069L,PhysRevB.93.075155,2011arXiv1105.4675W,resonancekfese}}
We have performed energy scans at the resonance wave vector of {\new $\bm{Q}=(0.25,\,0.75,\,0)$} at various temperatures. In Fig.~\ref{ins}(c), we show two such scans, measured at temperatures above and below \tc. Apparently, there is some spectral weight enhancement around the resonance energy of $E_r\approx14.5$~meV at 5~K. We have performed $\bm{Q}$ scans across {\new(0.25,\,0.75,\,0)} along the [110] direction at a series of temperatures down to 5~K. The integrated intensities obtained from the fitting to these scans are plotted as a function of temperature in Fig.~\ref{ins}(d). It is clear that the intensity starts to rise around $T_c$, and increases like the superconducting order parameter. Such a temperature dependence is prototypical for a neutron-spin resonance mode.\cite{RevModPhys.87.855} These results clearly show that the antiferromagnetic and superconducting orders both exist in this sample, as is commonly observed.\cite{RevModPhys.85.849} Furthermore, there is a reasonably strong response from the superconducting part of the sample, which makes it possible to explore the connection between the two orders.

To document this further, we have carried out elastic neutron scattering measurements on the sample {\new in the temperature range around \tc}. In Fig.~\ref{scan}a and b, we plot scans through the magnetic peak (0.2,\,0.4,\,0.5) along both the [120] and [001] directions at 30 and 55~K. From these scans, we determine the position of the magnetic peak to be at (0.193,\,0.386,\,0.5), but since the incommensurability of 0.007~rlu is smaller than the $\bm{Q}$ resolution, we still label the peak as (0.2,\,0.4,\,0.5). The peak widths are resolution limited both in and out of plane, indicating that the magnetic order extends over at least 500~{\AA} in three dimensions. Upon cooling, the magnetic peak intensity increases but the peak width and position do not show any noticeable changes. We have fitted the scans using Gaussian functions, and the integrated intensities obtained from fits to the [120] scans are plotted in Fig.~\ref{order}. When changing the temperature, to make sure that the sample alignment did not change, we always performed accompanying scans through the nuclear Bragg peaks. Two such scans at 30 and 55~K are plotted in Fig.~\ref{scan}(c); as one can see, they are virtually identical. Since our sample is large, to make sure that it reached thermal equilibrium after changing temperature, we waited for a sufficiently long time before starting the scans (10~min/K near 42~K). We also repeated scans during warming and cooling cycles, which confirmed that the results were reproducible.  

One can see from Fig.~\ref{order} that with decreasing temperature, the magnetic peak intensity increases linearly as the magnetic order keeps developing, {\new following the trend at higher temperatures,\cite{2011arXiv1102.0830B,2011arXiv1102.2882Y}}. At 42~K, the growth rate of the peak intensity is reduced. Below 42~K, the order evolves smoothly across $T_c$. In refs~\cite{2011arXiv1102.0830B,2011arXiv1102.2882Y}, Bao and his colleagues have reported a weaker anomaly, but in a similar temperature range. {\new Compared with the BaFe$_2$As$_2$ case, where superconductivity and magnetic order are believed to coexist microscopically and compete with each other, and thus the magnetic order parameter shows a sharp down turn at \tc,\cite{PhysRevLett.103.087001,PhysRevB.83.172503,PhysRevLett.110.147003,PhysRevB.81.134512}} our results are markedly different: i), the margin of the suppression is smaller; ii), the kink occurs about 11~K above \tc. Considering the differences in the suppression, the causes are almost certainly different. 

We think that the results are an indication that although the magnetic and superconducting orders are connected in some way, they do not coexist on a microscopic scale as expected from the phase diagram of ref.~\cite{PhysRevB.93.075155}. Indeed, optical conductivity experiments demonstrate the low average carrier density and provide evidence for Josephson coupling among superconducting grains.\cite{PhysRevB.86.144530,2011arXiv1102.1381Y} By assuming the A$_x$Fe$_{2-y}$Se$_2$ samples to be phase separated, Jiang \et reproduced the reduction of the ordered moment around \tc.\cite{2011arXiv1111.1860J} They argued that the proximity effect between the neighboring superconducting layers that was large due to the relatively weak correlation and large interlayer hopping, was responsible for the suppression. {\new In our sample, it is possible that superconducting fluctuations that set in at higher temperatures impact local regions of the dominant $\sqrt{5}\times\sqrt{5}$ phase by proximity effect, while coherent Josephson coupling among these domains only occurs at $T_c$. Furthermore, although from our own susceptibility and resistivity measurements we do not observe any signatures of superconductivity above \tc, there are some reports that this system may have a \tc higher than 40~K.\cite{sr_3_01216,2010arXiv1012.5236F,2011arXiv1101.0789W} If such a superconducting phase is present in our large sample, this suppression is expected. However, any such phase must be present as only a miniscule fraction. The fact that the magnetic order continues to rise, though at a slower rate below 42~K suggests that there are regions of the antiferromagnetic phase that feel little impact of the superconductivity, also supporting the phase-separation picture.} Another issue that may be relevant to the observation is the complex phases of this system.\cite{PhysRevB.93.075155} Now, it is known that in addition to the insulating and superconducting phases, other phases such as the K$_2$Fe$_7$Se$_8$,\cite{nc_4_1897} and semiconducting KFe$_{1.5}$Se$_2$ phases may also be present.\cite{PhysRevLett.109.267003} These additional phases may have different characteristic temperatures, and the possible presence of these phases further complicates the interplay among them. However, we have no direct evidence for the presence of such minority phases in our single crystal samples.   

\section{Conclusions}
{\new To summarize, we have shown that in the \kfs samples that contain the superconducting phase, the block antiferromagnetic order coexists with the high-temperature superconductivity albeit in separate, interdigitated, regions of the sample. When the temperature is decreased towards \tc, there is a well-defined suppression of the intensity of the magnetic Bragg peak, but at a temperature 11~K above the \tc. This result may possibly be understood in terms of the proximity effect of the superconducting fluctuations on the antiferromagnetic order.}

\section{Acknowledgements}
We thank Fuchun Zhang, Dunghai Lee, Qiang-Hua Wang, Hai-Hu Wen, and Jian-Xin Li for stimulating discussions. Work at Nanjing University was supported by NSFC Nos. 11374143 and 11674157. Work at Brookhaven National Laboratory (BNL) was supported by the Office of Basic Energy Sciences, US Department of Energy under Contract DE-SC0012704. R.D.Z. and J.A.S. were supported by the Center for Emergent Superconductivity, an Energy Frontier Research Center, headquartered at BNL, funded by US Department of Energy, under Contract DE-2009-BNL-PM015. Work at Berkeley was supported by the same Office through Contract DE-AC02-05CH11231 within the Quantum Materials Program (KC2202). Research conducted at Oak Ridge National Laboratory's HFIR and SNS was sponsored by the Scientific User Facilities Division, Office of Basic Energy Sciences, US Department
of Energy.


\begin{thebibliography}{35}%
\makeatletter
\providecommand \@ifxundefined [1]{%
 \@ifx{#1\undefined}
}%
\providecommand \@ifnum [1]{%
 \ifnum #1\expandafter \@firstoftwo
 \else \expandafter \@secondoftwo
 \fi
}%
\providecommand \@ifx [1]{%
 \ifx #1\expandafter \@firstoftwo
 \else \expandafter \@secondoftwo
 \fi
}%
\providecommand \natexlab [1]{#1}%
\providecommand \enquote  [1]{``#1''}%
\providecommand \bibnamefont  [1]{#1}%
\providecommand \bibfnamefont [1]{#1}%
\providecommand \citenamefont [1]{#1}%
\providecommand \href@noop [0]{\@secondoftwo}%
\providecommand \href [0]{\begingroup \@sanitize@url \@href}%
\providecommand \@href[1]{\@@startlink{#1}\@@href}%
\providecommand \@@href[1]{\endgroup#1\@@endlink}%
\providecommand \@sanitize@url [0]{\catcode `\\12\catcode `\$12\catcode
  `\&12\catcode `\#12\catcode `\^12\catcode `\_12\catcode `\%12\relax}%
\providecommand \@@startlink[1]{}%
\providecommand \@@endlink[0]{}%
\providecommand \url  [0]{\begingroup\@sanitize@url \@url }%
\providecommand \@url [1]{\endgroup\@href {#1}{\urlprefix }}%
\providecommand \urlprefix  [0]{URL }%
\providecommand \Eprint [0]{\href }%
\providecommand \doibase [0]{http://dx.doi.org/}%
\providecommand \selectlanguage [0]{\@gobble}%
\providecommand \bibinfo  [0]{\@secondoftwo}%
\providecommand \bibfield  [0]{\@secondoftwo}%
\providecommand \translation [1]{[#1]}%
\providecommand \BibitemOpen [0]{}%
\providecommand \bibitemStop [0]{}%
\providecommand \bibitemNoStop [0]{.\EOS\space}%
\providecommand \EOS [0]{\spacefactor3000\relax}%
\providecommand \BibitemShut  [1]{\csname bibitem#1\endcsname}%
\let\auto@bib@innerbib\@empty
\bibitem [{\citenamefont {Guo}\ \emph {et~al.}(2010)\citenamefont {Guo},
  \citenamefont {Jin}, \citenamefont {Wang}, \citenamefont {Wang},
  \citenamefont {Zhu}, \citenamefont {Zhou}, \citenamefont {He},\ and\
  \citenamefont {Chen}}]{scinkfese}%
  \BibitemOpen
  \bibfield  {author} {\bibinfo {author} {\bibfnamefont {J.}~\bibnamefont
  {Guo}}, \bibinfo {author} {\bibfnamefont {S.}~\bibnamefont {Jin}}, \bibinfo
  {author} {\bibfnamefont {G.}~\bibnamefont {Wang}}, \bibinfo {author}
  {\bibfnamefont {S.}~\bibnamefont {Wang}}, \bibinfo {author} {\bibfnamefont
  {K.}~\bibnamefont {Zhu}}, \bibinfo {author} {\bibfnamefont {T.}~\bibnamefont
  {Zhou}}, \bibinfo {author} {\bibfnamefont {M.}~\bibnamefont {He}}, \ and\
  \bibinfo {author} {\bibfnamefont {X.}~\bibnamefont {Chen}},\ }\href@noop {}
  {\bibfield  {journal} {\bibinfo  {journal} {Phys. Rev. B}\ }\textbf {\bibinfo
  {volume} {82}},\ \bibinfo {pages} {180520} (\bibinfo {year}
  {2010})}\BibitemShut {NoStop}%
\bibitem [{\citenamefont {Dagotto}(2013)}]{RevModPhys.85.849}%
  \BibitemOpen
  \bibfield  {author} {\bibinfo {author} {\bibfnamefont {E.}~\bibnamefont
  {Dagotto}},\ }\href {\doibase 10.1103/RevModPhys.85.849} {\bibfield
  {journal} {\bibinfo  {journal} {Rev. Mod. Phys.}\ }\textbf {\bibinfo {volume}
  {85}},\ \bibinfo {pages} {849} (\bibinfo {year} {2013})}\BibitemShut
  {NoStop}%
\bibitem [{\citenamefont {Ding}\ \emph {et~al.}(2008)\citenamefont {Ding},
  \citenamefont {Richard}, \citenamefont {Nakayama}, \citenamefont {Sugawara},
  \citenamefont {Arakane}, \citenamefont {Sekiba}, \citenamefont {Takayama},
  \citenamefont {Souma}, \citenamefont {Sato}, \citenamefont {Takahashi},
  \citenamefont {Wang}, \citenamefont {Dai}, \citenamefont {Fang},
  \citenamefont {Chen}, \citenamefont {Luo},\ and\ \citenamefont
  {Wang}}]{dinghongepl}%
  \BibitemOpen
  \bibfield  {author} {\bibinfo {author} {\bibfnamefont {H.}~\bibnamefont
  {Ding}}, \bibinfo {author} {\bibfnamefont {P.}~\bibnamefont {Richard}},
  \bibinfo {author} {\bibfnamefont {K.}~\bibnamefont {Nakayama}}, \bibinfo
  {author} {\bibfnamefont {K.}~\bibnamefont {Sugawara}}, \bibinfo {author}
  {\bibfnamefont {T.}~\bibnamefont {Arakane}}, \bibinfo {author} {\bibfnamefont
  {Y.}~\bibnamefont {Sekiba}}, \bibinfo {author} {\bibfnamefont
  {A.}~\bibnamefont {Takayama}}, \bibinfo {author} {\bibfnamefont
  {S.}~\bibnamefont {Souma}}, \bibinfo {author} {\bibfnamefont
  {T.}~\bibnamefont {Sato}}, \bibinfo {author} {\bibfnamefont {T.}~\bibnamefont
  {Takahashi}}, \bibinfo {author} {\bibfnamefont {Z.}~\bibnamefont {Wang}},
  \bibinfo {author} {\bibfnamefont {X.}~\bibnamefont {Dai}}, \bibinfo {author}
  {\bibfnamefont {Z.}~\bibnamefont {Fang}}, \bibinfo {author} {\bibfnamefont
  {G.~F.}\ \bibnamefont {Chen}}, \bibinfo {author} {\bibfnamefont {J.~L.}\
  \bibnamefont {Luo}}, \ and\ \bibinfo {author} {\bibfnamefont {N.~L.}\
  \bibnamefont {Wang}},\ }\href@noop {} {\bibfield  {journal} {\bibinfo
  {journal} {Euro. Phys. Lett.}\ }\textbf {\bibinfo {volume} {83}},\ \bibinfo
  {pages} {47001} (\bibinfo {year} {2008})}\BibitemShut {NoStop}%
\bibitem [{\citenamefont {Mou}\ \emph {et~al.}(2011)\citenamefont {Mou},
  \citenamefont {Liu}, \citenamefont {Jia}, \citenamefont {He}, \citenamefont
  {Peng}, \citenamefont {Zhao}, \citenamefont {Yu}, \citenamefont {Liu},
  \citenamefont {He}, \citenamefont {Dong}, \citenamefont {Zhang},
  \citenamefont {Wang}, \citenamefont {Dong}, \citenamefont {Fang},
  \citenamefont {Wang}, \citenamefont {Peng}, \citenamefont {Wang},
  \citenamefont {Zhang}, \citenamefont {Yang}, \citenamefont {Xu},
  \citenamefont {Chen},\ and\ \citenamefont {Zhou}}]{2011arXiv1101.4556M}%
  \BibitemOpen
  \bibfield  {author} {\bibinfo {author} {\bibfnamefont {D.}~\bibnamefont
  {Mou}}, \bibinfo {author} {\bibfnamefont {S.}~\bibnamefont {Liu}}, \bibinfo
  {author} {\bibfnamefont {X.}~\bibnamefont {Jia}}, \bibinfo {author}
  {\bibfnamefont {J.}~\bibnamefont {He}}, \bibinfo {author} {\bibfnamefont
  {Y.}~\bibnamefont {Peng}}, \bibinfo {author} {\bibfnamefont {L.}~\bibnamefont
  {Zhao}}, \bibinfo {author} {\bibfnamefont {L.}~\bibnamefont {Yu}}, \bibinfo
  {author} {\bibfnamefont {G.}~\bibnamefont {Liu}}, \bibinfo {author}
  {\bibfnamefont {S.}~\bibnamefont {He}}, \bibinfo {author} {\bibfnamefont
  {X.}~\bibnamefont {Dong}}, \bibinfo {author} {\bibfnamefont {J.}~\bibnamefont
  {Zhang}}, \bibinfo {author} {\bibfnamefont {H.}~\bibnamefont {Wang}},
  \bibinfo {author} {\bibfnamefont {C.}~\bibnamefont {Dong}}, \bibinfo {author}
  {\bibfnamefont {M.}~\bibnamefont {Fang}}, \bibinfo {author} {\bibfnamefont
  {X.}~\bibnamefont {Wang}}, \bibinfo {author} {\bibfnamefont {Q.}~\bibnamefont
  {Peng}}, \bibinfo {author} {\bibfnamefont {Z.}~\bibnamefont {Wang}}, \bibinfo
  {author} {\bibfnamefont {S.}~\bibnamefont {Zhang}}, \bibinfo {author}
  {\bibfnamefont {F.}~\bibnamefont {Yang}}, \bibinfo {author} {\bibfnamefont
  {Z.}~\bibnamefont {Xu}}, \bibinfo {author} {\bibfnamefont {C.}~\bibnamefont
  {Chen}}, \ and\ \bibinfo {author} {\bibfnamefont {X.~J.}\ \bibnamefont
  {Zhou}},\ }\href@noop {} {\bibfield  {journal} {\bibinfo  {journal} {Phys.
  Rev. Lett.}\ }\textbf {\bibinfo {volume} {106}},\ \bibinfo {pages} {107001}
  (\bibinfo {year} {2011})}\BibitemShut {NoStop}%
\bibitem [{\citenamefont {Wang}\ \emph
  {et~al.}(2011{\natexlab{a}})\citenamefont {Wang}, \citenamefont {Qian},
  \citenamefont {Richard}, \citenamefont {Zhang}, \citenamefont {Dong},
  \citenamefont {Wang}, \citenamefont {Dong}, \citenamefont {Fang},\ and\
  \citenamefont {Ding}}]{hongdingarpeskfese}%
  \BibitemOpen
  \bibfield  {author} {\bibinfo {author} {\bibfnamefont {X.-P.}\ \bibnamefont
  {Wang}}, \bibinfo {author} {\bibfnamefont {T.}~\bibnamefont {Qian}}, \bibinfo
  {author} {\bibfnamefont {P.}~\bibnamefont {Richard}}, \bibinfo {author}
  {\bibfnamefont {P.}~\bibnamefont {Zhang}}, \bibinfo {author} {\bibfnamefont
  {J.}~\bibnamefont {Dong}}, \bibinfo {author} {\bibfnamefont {H.-D.}\
  \bibnamefont {Wang}}, \bibinfo {author} {\bibfnamefont {C.-H.}\ \bibnamefont
  {Dong}}, \bibinfo {author} {\bibfnamefont {M.-H.}\ \bibnamefont {Fang}}, \
  and\ \bibinfo {author} {\bibfnamefont {H.}~\bibnamefont {Ding}},\ }\href@noop
  {} {\bibfield  {journal} {\bibinfo  {journal} {Euro. Phys. Lett.}\ }\textbf
  {\bibinfo {volume} {93}},\ \bibinfo {pages} {57001} (\bibinfo {year}
  {2011}{\natexlab{a}})}\BibitemShut {NoStop}%
\bibitem [{\citenamefont {{Zhang}}\ \emph {et~al.}(2011)\citenamefont
  {{Zhang}}, \citenamefont {{Yang}}, \citenamefont {{Xu}}, \citenamefont
  {{Ye}}, \citenamefont {{Chen}}, \citenamefont {{He}}, \citenamefont
  {{Jiang}}, \citenamefont {{Xie}}, \citenamefont {{Ying}}, \citenamefont
  {{Wang}}, \citenamefont {{Chen}}, \citenamefont {{Hu}},\ and\ \citenamefont
  {{Feng}}}]{kfesearpes1}%
  \BibitemOpen
  \bibfield  {author} {\bibinfo {author} {\bibfnamefont {Y.}~\bibnamefont
  {{Zhang}}}, \bibinfo {author} {\bibfnamefont {L.~X.}\ \bibnamefont {{Yang}}},
  \bibinfo {author} {\bibfnamefont {M.}~\bibnamefont {{Xu}}}, \bibinfo {author}
  {\bibfnamefont {Z.~R.}\ \bibnamefont {{Ye}}}, \bibinfo {author}
  {\bibfnamefont {F.}~\bibnamefont {{Chen}}}, \bibinfo {author} {\bibfnamefont
  {C.}~\bibnamefont {{He}}}, \bibinfo {author} {\bibfnamefont {J.}~\bibnamefont
  {{Jiang}}}, \bibinfo {author} {\bibfnamefont {B.~P.}\ \bibnamefont {{Xie}}},
  \bibinfo {author} {\bibfnamefont {J.~J.}\ \bibnamefont {{Ying}}}, \bibinfo
  {author} {\bibfnamefont {X.~F.}\ \bibnamefont {{Wang}}}, \bibinfo {author}
  {\bibfnamefont {X.~H.}\ \bibnamefont {{Chen}}}, \bibinfo {author}
  {\bibfnamefont {J.~P.}\ \bibnamefont {{Hu}}}, \ and\ \bibinfo {author}
  {\bibfnamefont {D.~L.}\ \bibnamefont {{Feng}}},\ }\href@noop {} {\bibfield
  {journal} {\bibinfo  {journal} {Nature Mater.}\ }\textbf {\bibinfo {volume}
  {10}},\ \bibinfo {pages} {273} (\bibinfo {year} {2011})}\BibitemShut
  {NoStop}%
\bibitem [{\citenamefont {Mazin}(2011)}]{Physics.4.26}%
  \BibitemOpen
  \bibfield  {author} {\bibinfo {author} {\bibfnamefont {I.}~\bibnamefont
  {Mazin}},\ }\href {\doibase 10.1103/Physics.4.26} {\bibfield  {journal}
  {\bibinfo  {journal} {Physics}\ }\textbf {\bibinfo {volume} {4}},\ \bibinfo
  {eid} {26} (\bibinfo {year} {2011})}\BibitemShut {NoStop}%
\bibitem [{\citenamefont {{Bao}}\ \emph {et~al.}(2011)\citenamefont {{Bao}},
  \citenamefont {{Huang}}, \citenamefont {{Chen}}, \citenamefont {{Green}},
  \citenamefont {{Wang}}, \citenamefont {{He}}, \citenamefont {{Wang}},\ and\
  \citenamefont {{Qiu}}}]{2011arXiv1102.0830B}%
  \BibitemOpen
  \bibfield  {author} {\bibinfo {author} {\bibfnamefont {W.}~\bibnamefont
  {{Bao}}}, \bibinfo {author} {\bibfnamefont {Q.}~\bibnamefont {{Huang}}},
  \bibinfo {author} {\bibfnamefont {G.~F.}\ \bibnamefont {{Chen}}}, \bibinfo
  {author} {\bibfnamefont {M.~A.}\ \bibnamefont {{Green}}}, \bibinfo {author}
  {\bibfnamefont {D.~M.}\ \bibnamefont {{Wang}}}, \bibinfo {author}
  {\bibfnamefont {J.~B.}\ \bibnamefont {{He}}}, \bibinfo {author}
  {\bibfnamefont {X.~Q.}\ \bibnamefont {{Wang}}}, \ and\ \bibinfo {author}
  {\bibfnamefont {Y.}~\bibnamefont {{Qiu}}},\ }\href@noop {} {\bibfield
  {journal} {\bibinfo  {journal} {Chin. Phys. Lett.}\ }\textbf {\bibinfo
  {volume} {28}},\ \bibinfo {pages} {086104} (\bibinfo {year}
  {2011})}\BibitemShut {NoStop}%
\bibitem [{\citenamefont {Ye}\ \emph {et~al.}(2011)\citenamefont {Ye},
  \citenamefont {Chi}, \citenamefont {Bao}, \citenamefont {Wang}, \citenamefont
  {Ying}, \citenamefont {Chen}, \citenamefont {Wang}, \citenamefont {Dong},\
  and\ \citenamefont {Fang}}]{2011arXiv1102.2882Y}%
  \BibitemOpen
  \bibfield  {author} {\bibinfo {author} {\bibfnamefont {F.}~\bibnamefont
  {Ye}}, \bibinfo {author} {\bibfnamefont {S.}~\bibnamefont {Chi}}, \bibinfo
  {author} {\bibfnamefont {W.}~\bibnamefont {Bao}}, \bibinfo {author}
  {\bibfnamefont {X.~F.}\ \bibnamefont {Wang}}, \bibinfo {author}
  {\bibfnamefont {J.~J.}\ \bibnamefont {Ying}}, \bibinfo {author}
  {\bibfnamefont {X.~H.}\ \bibnamefont {Chen}}, \bibinfo {author}
  {\bibfnamefont {H.~D.}\ \bibnamefont {Wang}}, \bibinfo {author}
  {\bibfnamefont {C.~H.}\ \bibnamefont {Dong}}, \ and\ \bibinfo {author}
  {\bibfnamefont {M.}~\bibnamefont {Fang}},\ }\href {\doibase
  10.1103/PhysRevLett.107.137003} {\bibfield  {journal} {\bibinfo  {journal}
  {Phys. Rev. Lett.}\ }\textbf {\bibinfo {volume} {107}},\ \bibinfo {pages}
  {137003} (\bibinfo {year} {2011})}\BibitemShut {NoStop}%
\bibitem [{\citenamefont {BAO~Wei}(2013)}]{BAOWei:027402}%
  \BibitemOpen
  \bibfield  {author} {\bibinfo {author} {\bibfnamefont {H.~Q.-Z. C. G.-F. H.
  J.-B. W. D.-M. M. A. G. Q. Y.-M. L. J.-L. W. M.-M.}\ \bibnamefont {BAO~Wei},
  \bibfnamefont {LI~Guan-Nan}},\ }\href {\doibase
  10.1088/0256-307X/30/2/027402} {\bibfield  {journal} {\bibinfo  {journal}
  {Chin. Phys. Lett.}\ }\textbf {\bibinfo {volume} {30}},\ \bibinfo {eid}
  {27402} (\bibinfo {year} {2013})}\BibitemShut {NoStop}%
\bibitem [{\citenamefont {Shermadini}\ \emph {et~al.}(2011)\citenamefont
  {Shermadini}, \citenamefont {Krzton-Maziopa}, \citenamefont {Bendele},
  \citenamefont {Khasanov}, \citenamefont {Luetkens}, \citenamefont {Conder},
  \citenamefont {Pomjakushina}, \citenamefont {Weyeneth}, \citenamefont
  {Pomjakushin}, \citenamefont {Bossen},\ and\ \citenamefont
  {Amato}}]{2011arXiv1101.1873S}%
  \BibitemOpen
  \bibfield  {author} {\bibinfo {author} {\bibfnamefont {Z.}~\bibnamefont
  {Shermadini}}, \bibinfo {author} {\bibfnamefont {A.}~\bibnamefont
  {Krzton-Maziopa}}, \bibinfo {author} {\bibfnamefont {M.}~\bibnamefont
  {Bendele}}, \bibinfo {author} {\bibfnamefont {R.}~\bibnamefont {Khasanov}},
  \bibinfo {author} {\bibfnamefont {H.}~\bibnamefont {Luetkens}}, \bibinfo
  {author} {\bibfnamefont {K.}~\bibnamefont {Conder}}, \bibinfo {author}
  {\bibfnamefont {E.}~\bibnamefont {Pomjakushina}}, \bibinfo {author}
  {\bibfnamefont {S.}~\bibnamefont {Weyeneth}}, \bibinfo {author}
  {\bibfnamefont {V.}~\bibnamefont {Pomjakushin}}, \bibinfo {author}
  {\bibfnamefont {O.}~\bibnamefont {Bossen}}, \ and\ \bibinfo {author}
  {\bibfnamefont {A.}~\bibnamefont {Amato}},\ }\href@noop {} {\bibfield
  {journal} {\bibinfo  {journal} {Phys. Rev. Lett.}\ }\textbf {\bibinfo
  {volume} {106}},\ \bibinfo {pages} {117602} (\bibinfo {year}
  {2011})}\BibitemShut {NoStop}%
\bibitem [{\citenamefont {Wang}\ \emph
  {et~al.}(2011{\natexlab{b}})\citenamefont {Wang}, \citenamefont {Song},
  \citenamefont {Shi}, \citenamefont {Wang}, \citenamefont {Chen},
  \citenamefont {Tian}, \citenamefont {Chen}, \citenamefont {Guo},
  \citenamefont {Yang},\ and\ \citenamefont {Li}}]{PhysRevB.83.140505}%
  \BibitemOpen
  \bibfield  {author} {\bibinfo {author} {\bibfnamefont {Z.}~\bibnamefont
  {Wang}}, \bibinfo {author} {\bibfnamefont {Y.~J.}\ \bibnamefont {Song}},
  \bibinfo {author} {\bibfnamefont {H.~L.}\ \bibnamefont {Shi}}, \bibinfo
  {author} {\bibfnamefont {Z.~W.}\ \bibnamefont {Wang}}, \bibinfo {author}
  {\bibfnamefont {Z.}~\bibnamefont {Chen}}, \bibinfo {author} {\bibfnamefont
  {H.~F.}\ \bibnamefont {Tian}}, \bibinfo {author} {\bibfnamefont {G.~F.}\
  \bibnamefont {Chen}}, \bibinfo {author} {\bibfnamefont {J.~G.}\ \bibnamefont
  {Guo}}, \bibinfo {author} {\bibfnamefont {H.~X.}\ \bibnamefont {Yang}}, \
  and\ \bibinfo {author} {\bibfnamefont {J.~Q.}\ \bibnamefont {Li}},\
  }\href@noop {} {\bibfield  {journal} {\bibinfo  {journal} {Phys. Rev. B}\
  }\textbf {\bibinfo {volume} {83}},\ \bibinfo {pages} {140505} (\bibinfo
  {year} {2011}{\natexlab{b}})}\BibitemShut {NoStop}%
\bibitem [{\citenamefont {Chen}\ \emph {et~al.}(2011)\citenamefont {Chen},
  \citenamefont {Xu}, \citenamefont {Ge}, \citenamefont {Zhang}, \citenamefont
  {Ye}, \citenamefont {Yang}, \citenamefont {Jiang}, \citenamefont {Xie},
  \citenamefont {Che}, \citenamefont {Zhang}, \citenamefont {Wang},
  \citenamefont {Chen}, \citenamefont {Shen}, \citenamefont {Hu},\ and\
  \citenamefont {Feng}}]{2011arXiv1106.3026C}%
  \BibitemOpen
  \bibfield  {author} {\bibinfo {author} {\bibfnamefont {F.}~\bibnamefont
  {Chen}}, \bibinfo {author} {\bibfnamefont {M.}~\bibnamefont {Xu}}, \bibinfo
  {author} {\bibfnamefont {Q.~Q.}\ \bibnamefont {Ge}}, \bibinfo {author}
  {\bibfnamefont {Y.}~\bibnamefont {Zhang}}, \bibinfo {author} {\bibfnamefont
  {Z.~R.}\ \bibnamefont {Ye}}, \bibinfo {author} {\bibfnamefont {L.~X.}\
  \bibnamefont {Yang}}, \bibinfo {author} {\bibfnamefont {J.}~\bibnamefont
  {Jiang}}, \bibinfo {author} {\bibfnamefont {B.~P.}\ \bibnamefont {Xie}},
  \bibinfo {author} {\bibfnamefont {R.~C.}\ \bibnamefont {Che}}, \bibinfo
  {author} {\bibfnamefont {M.}~\bibnamefont {Zhang}}, \bibinfo {author}
  {\bibfnamefont {A.~F.}\ \bibnamefont {Wang}}, \bibinfo {author}
  {\bibfnamefont {X.~H.}\ \bibnamefont {Chen}}, \bibinfo {author}
  {\bibfnamefont {D.~W.}\ \bibnamefont {Shen}}, \bibinfo {author}
  {\bibfnamefont {J.~P.}\ \bibnamefont {Hu}}, \ and\ \bibinfo {author}
  {\bibfnamefont {D.~L.}\ \bibnamefont {Feng}},\ }\href {\doibase
  10.1103/PhysRevX.1.021020} {\bibfield  {journal} {\bibinfo  {journal} {Phys.
  Rev. X}\ }\textbf {\bibinfo {volume} {1}},\ \bibinfo {pages} {021020}
  (\bibinfo {year} {2011})}\BibitemShut {NoStop}%
\bibitem [{\citenamefont {Ding}\ \emph {et~al.}(2013)\citenamefont {Ding},
  \citenamefont {Fang}, \citenamefont {Wang}, \citenamefont {Yang},
  \citenamefont {Liu}, \citenamefont {Deng}, \citenamefont {Ma}, \citenamefont
  {Meng}, \citenamefont {Hu},\ and\ \citenamefont {Wen}}]{nc_4_1897}%
  \BibitemOpen
  \bibfield  {author} {\bibinfo {author} {\bibfnamefont {X.}~\bibnamefont
  {Ding}}, \bibinfo {author} {\bibfnamefont {D.}~\bibnamefont {Fang}}, \bibinfo
  {author} {\bibfnamefont {Z.}~\bibnamefont {Wang}}, \bibinfo {author}
  {\bibfnamefont {H.}~\bibnamefont {Yang}}, \bibinfo {author} {\bibfnamefont
  {J.}~\bibnamefont {Liu}}, \bibinfo {author} {\bibfnamefont {Q.}~\bibnamefont
  {Deng}}, \bibinfo {author} {\bibfnamefont {G.}~\bibnamefont {Ma}}, \bibinfo
  {author} {\bibfnamefont {C.}~\bibnamefont {Meng}}, \bibinfo {author}
  {\bibfnamefont {Y.}~\bibnamefont {Hu}}, \ and\ \bibinfo {author}
  {\bibfnamefont {H.-H.}\ \bibnamefont {Wen}},\ }\href@noop {} {\bibfield
  {journal} {\bibinfo  {journal} {Nature Commun.}\ }\textbf {\bibinfo {volume}
  {4}},\ \bibinfo {pages} {1897} (\bibinfo {year} {2013})}\BibitemShut
  {NoStop}%
\bibitem [{\citenamefont {Ricci}\ \emph
  {et~al.}(2011{\natexlab{a}})\citenamefont {Ricci}, \citenamefont {Poccia},
  \citenamefont {Campi}, \citenamefont {Joseph}, \citenamefont {Arrighetti},
  \citenamefont {Barba}, \citenamefont {Reynolds}, \citenamefont {Burghammer},
  \citenamefont {Takeya}, \citenamefont {Mizuguchi}, \citenamefont {Takano},
  \citenamefont {Colapietro}, \citenamefont {Saini},\ and\ \citenamefont
  {Bianconi}}]{2011arXiv1107.0412R}%
  \BibitemOpen
  \bibfield  {author} {\bibinfo {author} {\bibfnamefont {A.}~\bibnamefont
  {Ricci}}, \bibinfo {author} {\bibfnamefont {N.}~\bibnamefont {Poccia}},
  \bibinfo {author} {\bibfnamefont {G.}~\bibnamefont {Campi}}, \bibinfo
  {author} {\bibfnamefont {B.}~\bibnamefont {Joseph}}, \bibinfo {author}
  {\bibfnamefont {G.}~\bibnamefont {Arrighetti}}, \bibinfo {author}
  {\bibfnamefont {L.}~\bibnamefont {Barba}}, \bibinfo {author} {\bibfnamefont
  {M.}~\bibnamefont {Reynolds}}, \bibinfo {author} {\bibfnamefont
  {M.}~\bibnamefont {Burghammer}}, \bibinfo {author} {\bibfnamefont
  {H.}~\bibnamefont {Takeya}}, \bibinfo {author} {\bibfnamefont
  {Y.}~\bibnamefont {Mizuguchi}}, \bibinfo {author} {\bibfnamefont
  {Y.}~\bibnamefont {Takano}}, \bibinfo {author} {\bibfnamefont
  {M.}~\bibnamefont {Colapietro}}, \bibinfo {author} {\bibfnamefont {N.~L.}\
  \bibnamefont {Saini}}, \ and\ \bibinfo {author} {\bibfnamefont
  {A.}~\bibnamefont {Bianconi}},\ }\href {\doibase 10.1103/PhysRevB.84.060511}
  {\bibfield  {journal} {\bibinfo  {journal} {Phys. Rev. B}\ }\textbf {\bibinfo
  {volume} {84}},\ \bibinfo {pages} {060511} (\bibinfo {year}
  {2011}{\natexlab{a}})}\BibitemShut {NoStop}%
\bibitem [{\citenamefont {Ricci}\ \emph
  {et~al.}(2011{\natexlab{b}})\citenamefont {Ricci}, \citenamefont {Poccia},
  \citenamefont {Joseph}, \citenamefont {Arrighetti}, \citenamefont {Barba},
  \citenamefont {Plaisier}, \citenamefont {Campi}, \citenamefont {Mizuguchi},
  \citenamefont {Takeya}, \citenamefont {Takano}, \citenamefont {Saini},\ and\
  \citenamefont {Bianconi}}]{phasesep122_1}%
  \BibitemOpen
  \bibfield  {author} {\bibinfo {author} {\bibfnamefont {A.}~\bibnamefont
  {Ricci}}, \bibinfo {author} {\bibfnamefont {N.}~\bibnamefont {Poccia}},
  \bibinfo {author} {\bibfnamefont {B.}~\bibnamefont {Joseph}}, \bibinfo
  {author} {\bibfnamefont {G.}~\bibnamefont {Arrighetti}}, \bibinfo {author}
  {\bibfnamefont {L.}~\bibnamefont {Barba}}, \bibinfo {author} {\bibfnamefont
  {J.}~\bibnamefont {Plaisier}}, \bibinfo {author} {\bibfnamefont
  {G.}~\bibnamefont {Campi}}, \bibinfo {author} {\bibfnamefont
  {Y.}~\bibnamefont {Mizuguchi}}, \bibinfo {author} {\bibfnamefont
  {H.}~\bibnamefont {Takeya}}, \bibinfo {author} {\bibfnamefont
  {Y.}~\bibnamefont {Takano}}, \bibinfo {author} {\bibfnamefont {N.~L.}\
  \bibnamefont {Saini}}, \ and\ \bibinfo {author} {\bibfnamefont
  {A.}~\bibnamefont {Bianconi}},\ }\href@noop {} {\bibfield  {journal}
  {\bibinfo  {journal} {Supercond. Sci. Tech.}\ }\textbf {\bibinfo {volume}
  {24}},\ \bibinfo {pages} {082002} (\bibinfo {year}
  {2011}{\natexlab{b}})}\BibitemShut {NoStop}%
\bibitem [{\citenamefont {Li}\ \emph {et~al.}(2012)\citenamefont {Li},
  \citenamefont {Ding}, \citenamefont {Deng}, \citenamefont {Chang},
  \citenamefont {Song}, \citenamefont {He}, \citenamefont {Wang}, \citenamefont
  {Ma}, \citenamefont {Hu}, \citenamefont {Chen},\ and\ \citenamefont
  {Xue}}]{2011arXiv1108.0069L}%
  \BibitemOpen
  \bibfield  {author} {\bibinfo {author} {\bibfnamefont {W.}~\bibnamefont
  {Li}}, \bibinfo {author} {\bibfnamefont {H.}~\bibnamefont {Ding}}, \bibinfo
  {author} {\bibfnamefont {P.}~\bibnamefont {Deng}}, \bibinfo {author}
  {\bibfnamefont {K.}~\bibnamefont {Chang}}, \bibinfo {author} {\bibfnamefont
  {C.}~\bibnamefont {Song}}, \bibinfo {author} {\bibfnamefont {K.}~\bibnamefont
  {He}}, \bibinfo {author} {\bibfnamefont {L.}~\bibnamefont {Wang}}, \bibinfo
  {author} {\bibfnamefont {X.}~\bibnamefont {Ma}}, \bibinfo {author}
  {\bibfnamefont {J.-P.}\ \bibnamefont {Hu}}, \bibinfo {author} {\bibfnamefont
  {X.}~\bibnamefont {Chen}}, \ and\ \bibinfo {author} {\bibfnamefont {Q.-K.}\
  \bibnamefont {Xue}},\ }\href@noop {} {\bibfield  {journal} {\bibinfo
  {journal} {Nature Phys.}\ }\textbf {\bibinfo {volume} {8}},\ \bibinfo {pages}
  {126} (\bibinfo {year} {2012})}.%
\bibitem [{\citenamefont {Wang}\ \emph {et~al.}(2016)\citenamefont {Wang},
  \citenamefont {Yi}, \citenamefont {Tian}, \citenamefont
  {Bourret-Courchesne},\ and\ \citenamefont {Birgeneau}}]{PhysRevB.93.075155}%
  \BibitemOpen
  \bibfield  {author} {\bibinfo {author} {\bibfnamefont {M.}~\bibnamefont
  {Wang}}, \bibinfo {author} {\bibfnamefont {M.}~\bibnamefont {Yi}}, \bibinfo
  {author} {\bibfnamefont {W.}~\bibnamefont {Tian}}, \bibinfo {author}
  {\bibfnamefont {E.}~\bibnamefont {Bourret-Courchesne}}, \ and\ \bibinfo
  {author} {\bibfnamefont {R.~J.}\ \bibnamefont {Birgeneau}},\ }\href {\doibase
  10.1103/PhysRevB.93.075155} {\bibfield  {journal} {\bibinfo  {journal} {Phys.
  Rev. B}\ }\textbf {\bibinfo {volume} {93}},\ \bibinfo {pages} {075155}
  (\bibinfo {year} {2016})}\BibitemShut {NoStop}%
\bibitem [{\citenamefont {Pratt}\ \emph {et~al.}(2009)\citenamefont {Pratt},
  \citenamefont {Tian}, \citenamefont {Kreyssig}, \citenamefont {Zarestky},
  \citenamefont {Nandi}, \citenamefont {Ni}, \citenamefont {Bud'ko},
  \citenamefont {Canfield}, \citenamefont {Goldman},\ and\ \citenamefont
  {McQueeney}}]{PhysRevLett.103.087001}%
  \BibitemOpen
  \bibfield  {author} {\bibinfo {author} {\bibfnamefont {D.~K.}\ \bibnamefont
  {Pratt}}, \bibinfo {author} {\bibfnamefont {W.}~\bibnamefont {Tian}},
  \bibinfo {author} {\bibfnamefont {A.}~\bibnamefont {Kreyssig}}, \bibinfo
  {author} {\bibfnamefont {J.~L.}\ \bibnamefont {Zarestky}}, \bibinfo {author}
  {\bibfnamefont {S.}~\bibnamefont {Nandi}}, \bibinfo {author} {\bibfnamefont
  {N.}~\bibnamefont {Ni}}, \bibinfo {author} {\bibfnamefont {S.~L.}\
  \bibnamefont {Bud'ko}}, \bibinfo {author} {\bibfnamefont {P.~C.}\
  \bibnamefont {Canfield}}, \bibinfo {author} {\bibfnamefont {A.~I.}\
  \bibnamefont {Goldman}}, \ and\ \bibinfo {author} {\bibfnamefont {R.~J.}\
  \bibnamefont {McQueeney}},\ }\href@noop {} {\bibfield  {journal} {\bibinfo
  {journal} {Phys. Rev. Lett.}\ }\textbf {\bibinfo {volume} {103}},\ \bibinfo
  {pages} {087001} (\bibinfo {year} {2009})}\BibitemShut {NoStop}%
\bibitem [{\citenamefont {Avci}\ \emph {et~al.}(2011)\citenamefont {Avci},
  \citenamefont {Chmaissem}, \citenamefont {Goremychkin}, \citenamefont
  {Rosenkranz}, \citenamefont {Castellan}, \citenamefont {Chung}, \citenamefont
  {Todorov}, \citenamefont {Schlueter}, \citenamefont {Claus}, \citenamefont
  {Kanatzidis}, \citenamefont {Daoud-Aladine}, \citenamefont {Khalyavin},\ and\
  \citenamefont {Osborn}}]{PhysRevB.83.172503}%
  \BibitemOpen
  \bibfield  {author} {\bibinfo {author} {\bibfnamefont {S.}~\bibnamefont
  {Avci}}, \bibinfo {author} {\bibfnamefont {O.}~\bibnamefont {Chmaissem}},
  \bibinfo {author} {\bibfnamefont {E.~A.}\ \bibnamefont {Goremychkin}},
  \bibinfo {author} {\bibfnamefont {S.}~\bibnamefont {Rosenkranz}}, \bibinfo
  {author} {\bibfnamefont {J.-P.}\ \bibnamefont {Castellan}}, \bibinfo {author}
  {\bibfnamefont {D.~Y.}\ \bibnamefont {Chung}}, \bibinfo {author}
  {\bibfnamefont {I.~S.}\ \bibnamefont {Todorov}}, \bibinfo {author}
  {\bibfnamefont {J.~A.}\ \bibnamefont {Schlueter}}, \bibinfo {author}
  {\bibfnamefont {H.}~\bibnamefont {Claus}}, \bibinfo {author} {\bibfnamefont
  {M.~G.}\ \bibnamefont {Kanatzidis}}, \bibinfo {author} {\bibfnamefont
  {A.}~\bibnamefont {Daoud-Aladine}}, \bibinfo {author} {\bibfnamefont
  {D.}~\bibnamefont {Khalyavin}}, \ and\ \bibinfo {author} {\bibfnamefont
  {R.}~\bibnamefont {Osborn}},\ }\href {\doibase 10.1103/PhysRevB.83.172503}
  {\bibfield  {journal} {\bibinfo  {journal} {Phys. Rev. B}\ }\textbf {\bibinfo
  {volume} {83}},\ \bibinfo {pages} {172503} (\bibinfo {year}
  {2011})}\BibitemShut {NoStop}%
\bibitem [{\citenamefont {Zhao}\ \emph {et~al.}(2013)\citenamefont {Zhao},
  \citenamefont {Rotundu}, \citenamefont {Marty}, \citenamefont {Matsuda},
  \citenamefont {Zhao}, \citenamefont {Setty}, \citenamefont
  {Bourret-Courchesne}, \citenamefont {Hu},\ and\ \citenamefont
  {Birgeneau}}]{PhysRevLett.110.147003}%
  \BibitemOpen
  \bibfield  {author} {\bibinfo {author} {\bibfnamefont {J.}~\bibnamefont
  {Zhao}}, \bibinfo {author} {\bibfnamefont {C.~R.}\ \bibnamefont {Rotundu}},
  \bibinfo {author} {\bibfnamefont {K.}~\bibnamefont {Marty}}, \bibinfo
  {author} {\bibfnamefont {M.}~\bibnamefont {Matsuda}}, \bibinfo {author}
  {\bibfnamefont {Y.}~\bibnamefont {Zhao}}, \bibinfo {author} {\bibfnamefont
  {C.}~\bibnamefont {Setty}}, \bibinfo {author} {\bibfnamefont
  {E.}~\bibnamefont {Bourret-Courchesne}}, \bibinfo {author} {\bibfnamefont
  {J.}~\bibnamefont {Hu}}, \ and\ \bibinfo {author} {\bibfnamefont {R.~J.}\
  \bibnamefont {Birgeneau}},\ }\href {\doibase 10.1103/PhysRevLett.110.147003}
  {\bibfield  {journal} {\bibinfo  {journal} {Phys. Rev. Lett.}\ }\textbf
  {\bibinfo {volume} {110}},\ \bibinfo {pages} {147003} (\bibinfo {year}
  {2013})}\BibitemShut {NoStop}%
\bibitem [{\citenamefont {Kreyssig}\ \emph {et~al.}(2010)\citenamefont
  {Kreyssig}, \citenamefont {Kim}, \citenamefont {Nandi}, \citenamefont
  {Pratt}, \citenamefont {Tian}, \citenamefont {Zarestky}, \citenamefont {Ni},
  \citenamefont {Thaler}, \citenamefont {Bud'ko}, \citenamefont {Canfield},
  \citenamefont {McQueeney},\ and\ \citenamefont
  {Goldman}}]{PhysRevB.81.134512}%
  \BibitemOpen
  \bibfield  {author} {\bibinfo {author} {\bibfnamefont {A.}~\bibnamefont
  {Kreyssig}}, \bibinfo {author} {\bibfnamefont {M.~G.}\ \bibnamefont {Kim}},
  \bibinfo {author} {\bibfnamefont {S.}~\bibnamefont {Nandi}}, \bibinfo
  {author} {\bibfnamefont {D.~K.}\ \bibnamefont {Pratt}}, \bibinfo {author}
  {\bibfnamefont {W.}~\bibnamefont {Tian}}, \bibinfo {author} {\bibfnamefont
  {J.~L.}\ \bibnamefont {Zarestky}}, \bibinfo {author} {\bibfnamefont
  {N.}~\bibnamefont {Ni}}, \bibinfo {author} {\bibfnamefont {A.}~\bibnamefont
  {Thaler}}, \bibinfo {author} {\bibfnamefont {S.~L.}\ \bibnamefont {Bud'ko}},
  \bibinfo {author} {\bibfnamefont {P.~C.}\ \bibnamefont {Canfield}}, \bibinfo
  {author} {\bibfnamefont {R.~J.}\ \bibnamefont {McQueeney}}, \ and\ \bibinfo
  {author} {\bibfnamefont {A.~I.}\ \bibnamefont {Goldman}},\ }\href {\doibase
  10.1103/PhysRevB.81.134512} {\bibfield  {journal} {\bibinfo  {journal} {Phys.
  Rev. B}\ }\textbf {\bibinfo {volume} {81}},\ \bibinfo {pages} {134512}
  (\bibinfo {year} {2010})}\BibitemShut {NoStop}%
\bibitem [{\citenamefont {Zhao}\ \emph {et~al.}(2012)\citenamefont {Zhao},
  \citenamefont {Cao}, \citenamefont {Bourret-Courchesne}, \citenamefont
  {Lee},\ and\ \citenamefont {Birgeneau}}]{PhysRevLett.109.267003}%
  \BibitemOpen
  \bibfield  {author} {\bibinfo {author} {\bibfnamefont {J.}~\bibnamefont
  {Zhao}}, \bibinfo {author} {\bibfnamefont {H.}~\bibnamefont {Cao}}, \bibinfo
  {author} {\bibfnamefont {E.}~\bibnamefont {Bourret-Courchesne}}, \bibinfo
  {author} {\bibfnamefont {D.-H.}\ \bibnamefont {Lee}}, \ and\ \bibinfo
  {author} {\bibfnamefont {R.~J.}\ \bibnamefont {Birgeneau}},\ }\href {\doibase
  10.1103/PhysRevLett.109.267003} {\bibfield  {journal} {\bibinfo  {journal}
  {Phys. Rev. Lett.}\ }\textbf {\bibinfo {volume} {109}},\ \bibinfo {pages}
  {267003} (\bibinfo {year} {2012})}\BibitemShut {NoStop}%
\bibitem [{\citenamefont {Wen}\ \emph {et~al.}(2011)\citenamefont {Wen},
  \citenamefont {Xu}, \citenamefont {Gu}, \citenamefont {Tranquada},\ and\
  \citenamefont {Birgeneau}}]{interplaywen}%
  \BibitemOpen
  \bibfield  {author} {\bibinfo {author} {\bibfnamefont {J.}~\bibnamefont
  {Wen}}, \bibinfo {author} {\bibfnamefont {G.}~\bibnamefont {Xu}}, \bibinfo
  {author} {\bibfnamefont {G.}~\bibnamefont {Gu}}, \bibinfo {author}
  {\bibfnamefont {J.~M.}\ \bibnamefont {Tranquada}}, \ and\ \bibinfo {author}
  {\bibfnamefont {R.~J.}\ \bibnamefont {Birgeneau}},\ }\href@noop {} {\bibfield
   {journal} {\bibinfo  {journal} {Rep. Pro. Phys.}\ }\textbf {\bibinfo
  {volume} {74}},\ \bibinfo {pages} {124503} (\bibinfo {year}
  {2011})}\BibitemShut {NoStop}%
\bibitem [{\citenamefont {Homes}\ \emph
  {et~al.}(2012{\natexlab{a}})\citenamefont {Homes}, \citenamefont {Xu},
  \citenamefont {Wen},\ and\ \citenamefont {Gu}}]{PhysRevB.86.144530}%
  \BibitemOpen
  \bibfield  {author} {\bibinfo {author} {\bibfnamefont {C.~C.}\ \bibnamefont
  {Homes}}, \bibinfo {author} {\bibfnamefont {Z.~J.}\ \bibnamefont {Xu}},
  \bibinfo {author} {\bibfnamefont {J.~S.}\ \bibnamefont {Wen}}, \ and\
  \bibinfo {author} {\bibfnamefont {G.~D.}\ \bibnamefont {Gu}},\ }\href
  {\doibase 10.1103/PhysRevB.86.144530} {\bibfield  {journal} {\bibinfo
  {journal} {Phys. Rev. B}\ }\textbf {\bibinfo {volume} {86}},\ \bibinfo
  {pages} {144530} (\bibinfo {year} {2012}{\natexlab{a}})}\BibitemShut
  {NoStop}%
\bibitem [{\citenamefont {Tyson}\ \emph {et~al.}(2012)\citenamefont {Tyson},
  \citenamefont {Yu}, \citenamefont {Han}, \citenamefont {Croft}, \citenamefont
  {Gu}, \citenamefont {Dimitrov},\ and\ \citenamefont
  {Li}}]{2011arXiv1111.1026T}%
  \BibitemOpen
  \bibfield  {author} {\bibinfo {author} {\bibfnamefont {T.~A.}\ \bibnamefont
  {Tyson}}, \bibinfo {author} {\bibfnamefont {T.}~\bibnamefont {Yu}}, \bibinfo
  {author} {\bibfnamefont {S.~J.}\ \bibnamefont {Han}}, \bibinfo {author}
  {\bibfnamefont {M.}~\bibnamefont {Croft}}, \bibinfo {author} {\bibfnamefont
  {G.~D.}\ \bibnamefont {Gu}}, \bibinfo {author} {\bibfnamefont {I.~K.}\
  \bibnamefont {Dimitrov}}, \ and\ \bibinfo {author} {\bibfnamefont
  {Q.}~\bibnamefont {Li}},\ }\href {\doibase 10.1103/PhysRevB.85.024504}
  {\bibfield  {journal} {\bibinfo  {journal} {Phys. Rev. B}\ }\textbf {\bibinfo
  {volume} {85}},\ \bibinfo {pages} {024504} (\bibinfo {year}
  {2012})}\BibitemShut {NoStop}%
\bibitem [{\citenamefont {Homes}\ \emph
  {et~al.}(2012{\natexlab{b}})\citenamefont {Homes}, \citenamefont {Xu},
  \citenamefont {Wen},\ and\ \citenamefont {Gu}}]{2011arXiv1110.5529H}%
  \BibitemOpen
  \bibfield  {author} {\bibinfo {author} {\bibfnamefont {C.~C.}\ \bibnamefont
  {Homes}}, \bibinfo {author} {\bibfnamefont {Z.~J.}\ \bibnamefont {Xu}},
  \bibinfo {author} {\bibfnamefont {J.~S.}\ \bibnamefont {Wen}}, \ and\
  \bibinfo {author} {\bibfnamefont {G.~D.}\ \bibnamefont {Gu}},\ }\href
  {\doibase 10.1103/PhysRevB.85.180510} {\bibfield  {journal} {\bibinfo
  {journal} {Phys. Rev. B}\ }\textbf {\bibinfo {volume} {85}},\ \bibinfo
  {pages} {180510} (\bibinfo {year} {2012}{\natexlab{b}})}\BibitemShut
  {NoStop}%
\bibitem [{\citenamefont {Jiang}\ \emph {et~al.}(2012)\citenamefont {Jiang},
  \citenamefont {Chen}, \citenamefont {Yao},\ and\ \citenamefont
  {Zhang}}]{2011arXiv1111.1860J}%
  \BibitemOpen
  \bibfield  {author} {\bibinfo {author} {\bibfnamefont {H.-M.}\ \bibnamefont
  {Jiang}}, \bibinfo {author} {\bibfnamefont {W.-Q.}\ \bibnamefont {Chen}},
  \bibinfo {author} {\bibfnamefont {Z.-J.}\ \bibnamefont {Yao}}, \ and\
  \bibinfo {author} {\bibfnamefont {F.-C.}\ \bibnamefont {Zhang}},\ }\href
  {\doibase 10.1103/PhysRevB.85.104506} {\bibfield  {journal} {\bibinfo
  {journal} {Phys. Rev. B}\ }\textbf {\bibinfo {volume} {85}},\ \bibinfo
  {pages} {104506} (\bibinfo {year} {2012})}\BibitemShut {NoStop}%
\bibitem [{\citenamefont {{Wang}}\ \emph {et~al.}(2011)\citenamefont {{Wang}},
  \citenamefont {{Fang}}, \citenamefont {{Yao}}, \citenamefont {{Tan}},
  \citenamefont {{Harriger}}, \citenamefont {{Song}}, \citenamefont
  {{Netherton}}, \citenamefont {{Zhang}}, \citenamefont {{Wang}}, \citenamefont
  {{Stone}}, \citenamefont {{Tian}}, \citenamefont {{Hu}},\ and\ \citenamefont
  {{Dai}}}]{2011arXiv1105.4675W}%
  \BibitemOpen
  \bibfield  {author} {\bibinfo {author} {\bibfnamefont {M.}~\bibnamefont
  {{Wang}}}, \bibinfo {author} {\bibfnamefont {C.}~\bibnamefont {{Fang}}},
  \bibinfo {author} {\bibfnamefont {D.-X.}\ \bibnamefont {{Yao}}}, \bibinfo
  {author} {\bibfnamefont {G.}~\bibnamefont {{Tan}}}, \bibinfo {author}
  {\bibfnamefont {L.~W.}\ \bibnamefont {{Harriger}}}, \bibinfo {author}
  {\bibfnamefont {Y.}~\bibnamefont {{Song}}}, \bibinfo {author} {\bibfnamefont
  {T.}~\bibnamefont {{Netherton}}}, \bibinfo {author} {\bibfnamefont
  {C.}~\bibnamefont {{Zhang}}}, \bibinfo {author} {\bibfnamefont
  {M.}~\bibnamefont {{Wang}}}, \bibinfo {author} {\bibfnamefont {M.~B.}\
  \bibnamefont {{Stone}}}, \bibinfo {author} {\bibfnamefont {W.}~\bibnamefont
  {{Tian}}}, \bibinfo {author} {\bibfnamefont {J.}~\bibnamefont {{Hu}}}, \ and\
  \bibinfo {author} {\bibfnamefont {P.}~\bibnamefont {{Dai}}},\ }\href@noop {}
  {\bibfield  {journal} {\bibinfo  {journal} {Nature Commun.}\ }\textbf
  {\bibinfo {volume} {2}},\ \bibinfo {pages} {580} (\bibinfo {year}
  {2011})}\BibitemShut {NoStop}%
\bibitem [{\citenamefont {Park}\ \emph {et~al.}(2011)\citenamefont {Park},
  \citenamefont {Friemel}, \citenamefont {Li}, \citenamefont {Kim},
  \citenamefont {Tsurkan}, \citenamefont {Deisenhofer}, \citenamefont {Krug~von
  Nidda}, \citenamefont {Loidl}, \citenamefont {Ivanov}, \citenamefont
  {Keimer},\ and\ \citenamefont {Inosov}}]{resonancekfese}%
  \BibitemOpen
  \bibfield  {author} {\bibinfo {author} {\bibfnamefont {J.~T.}\ \bibnamefont
  {Park}}, \bibinfo {author} {\bibfnamefont {G.}~\bibnamefont {Friemel}},
  \bibinfo {author} {\bibfnamefont {Y.}~\bibnamefont {Li}}, \bibinfo {author}
  {\bibfnamefont {J.-H.}\ \bibnamefont {Kim}}, \bibinfo {author} {\bibfnamefont
  {V.}~\bibnamefont {Tsurkan}}, \bibinfo {author} {\bibfnamefont
  {J.}~\bibnamefont {Deisenhofer}}, \bibinfo {author} {\bibfnamefont {H.-A.}\
  \bibnamefont {Krug~von Nidda}}, \bibinfo {author} {\bibfnamefont
  {A.}~\bibnamefont {Loidl}}, \bibinfo {author} {\bibfnamefont
  {A.}~\bibnamefont {Ivanov}}, \bibinfo {author} {\bibfnamefont
  {B.}~\bibnamefont {Keimer}}, \ and\ \bibinfo {author} {\bibfnamefont {D.~S.}\
  \bibnamefont {Inosov}},\ }\href {\doibase 10.1103/PhysRevLett.107.177005}
  {\bibfield  {journal} {\bibinfo  {journal} {Phys. Rev. Lett.}\ }\textbf
  {\bibinfo {volume} {107}},\ \bibinfo {pages} {177005} (\bibinfo {year}
  {2011})}\BibitemShut {NoStop}%
\bibitem [{\citenamefont {Dai}(2015)}]{RevModPhys.87.855}%
  \BibitemOpen
  \bibfield  {author} {\bibinfo {author} {\bibfnamefont {P.}~\bibnamefont
  {Dai}},\ }\href {\doibase 10.1103/RevModPhys.87.855} {\bibfield  {journal}
  {\bibinfo  {journal} {Rev. Mod. Phys.}\ }\textbf {\bibinfo {volume} {87}},\
  \bibinfo {pages} {855} (\bibinfo {year} {2015})}\BibitemShut {NoStop}%
\bibitem [{\citenamefont {Yuan}\ \emph {et~al.}(2012)\citenamefont {Yuan},
  \citenamefont {Dong}, \citenamefont {Song}, \citenamefont {Zheng},
  \citenamefont {Chen}, \citenamefont {Hu}, \citenamefont {Li},\ and\
  \citenamefont {Wang}}]{2011arXiv1102.1381Y}%
  \BibitemOpen
  \bibfield  {author} {\bibinfo {author} {\bibfnamefont {R.~H.}\ \bibnamefont
  {Yuan}}, \bibinfo {author} {\bibfnamefont {T.}~\bibnamefont {Dong}}, \bibinfo
  {author} {\bibfnamefont {Y.~J.}\ \bibnamefont {Song}}, \bibinfo {author}
  {\bibfnamefont {P.}~\bibnamefont {Zheng}}, \bibinfo {author} {\bibfnamefont
  {G.~F.}\ \bibnamefont {Chen}}, \bibinfo {author} {\bibfnamefont {J.~P.}\
  \bibnamefont {Hu}}, \bibinfo {author} {\bibfnamefont {J.~Q.}\ \bibnamefont
  {Li}}, \ and\ \bibinfo {author} {\bibfnamefont {N.~L.}\ \bibnamefont
  {Wang}},\ }\href@noop {} {\bibfield  {journal} {\bibinfo  {journal}
  {Sci. Rep.}\ }\textbf {\bibinfo {volume} {2}},\ \bibinfo {pages}
  {221} (\bibinfo {year} {2012})}\BibitemShut {NoStop}%
\bibitem [{\citenamefont {Zhang}\ \emph {et~al.}(2013)\citenamefont {Zhang},
  \citenamefont {Xia}, \citenamefont {Liu}, \citenamefont {Tong}, \citenamefont
  {Yang},\ and\ \citenamefont {Zhang}}]{sr_3_01216}%
  \BibitemOpen
  \bibfield  {author} {\bibinfo {author} {\bibfnamefont {A.-m.}\ \bibnamefont
  {Zhang}}, \bibinfo {author} {\bibfnamefont {T.-l.}\ \bibnamefont {Xia}},
  \bibinfo {author} {\bibfnamefont {K.}~\bibnamefont {Liu}}, \bibinfo {author}
  {\bibfnamefont {W.}~\bibnamefont {Tong}}, \bibinfo {author} {\bibfnamefont
  {Z.-r.}\ \bibnamefont {Yang}}, \ and\ \bibinfo {author} {\bibfnamefont
  {Q.-m.}\ \bibnamefont {Zhang}},\ }\href@noop {} {\bibfield  {journal}
  {\bibinfo  {journal} {Sci. Rep.}\ } \textbf {\bibinfo {volume} {3}}, {\bibinfo  {pages} {1216}\ } (\bibinfo
  {year} {2013})}.%
\bibitem [{\citenamefont {Fang}\ \emph {et~al.}(2011)\citenamefont {Fang},
  \citenamefont {Wang}, \citenamefont {Dong}, \citenamefont {Li}, \citenamefont
  {Feng}, \citenamefont {Chen},\ and\ \citenamefont
  {Yuan}}]{2010arXiv1012.5236F}%
  \BibitemOpen
  \bibfield  {author} {\bibinfo {author} {\bibfnamefont {M.-H.}\ \bibnamefont
  {Fang}}, \bibinfo {author} {\bibfnamefont {H.-D.}\ \bibnamefont {Wang}},
  \bibinfo {author} {\bibfnamefont {C.-H.}\ \bibnamefont {Dong}}, \bibinfo
  {author} {\bibfnamefont {Z.-J.}\ \bibnamefont {Li}}, \bibinfo {author}
  {\bibfnamefont {C.-M.}\ \bibnamefont {Feng}}, \bibinfo {author}
  {\bibfnamefont {J.}~\bibnamefont {Chen}}, \ and\ \bibinfo {author}
  {\bibfnamefont {H.~Q.}\ \bibnamefont {Yuan}},\ }\href@noop {} {\bibfield
  {journal} {\bibinfo  {journal} {Euro. Phys. Lett.}\ }\textbf {\bibinfo
  {volume} {94}},\ \bibinfo {pages} {27009} (\bibinfo {year}
  {2011})}\BibitemShut {NoStop}%
\bibitem [{\citenamefont {Wang}\ \emph {et~al.}(2011)\citenamefont {Wang},
  \citenamefont {He}, \citenamefont {Xia},\ and\ \citenamefont
  {Chen}}]{2011arXiv1101.0789W}%
  \BibitemOpen
  \bibfield  {author} {\bibinfo {author} {\bibfnamefont {D.~M.}\ \bibnamefont
  {Wang}}, \bibinfo {author} {\bibfnamefont {J.~B.}\ \bibnamefont {He}},
  \bibinfo {author} {\bibfnamefont {T.-L.}\ \bibnamefont {Xia}}, \ and\
  \bibinfo {author} {\bibfnamefont {G.~F.}\ \bibnamefont {Chen}},\ }\href@noop
  {} {\bibfield  {journal} {\bibinfo  {journal} {Phys. Rev. B}\ }\textbf
  {\bibinfo {volume} {83}},\ \bibinfo {pages} {132502} (\bibinfo {year}
  {2011})}\BibitemShut {NoStop}%
\end{thebibliography}

%

\end{document}